\documentclass[aps, prab, reprint, showpacs, amsmath, amssymb]{revtex4-1}

\usepackage{amsmath,amscd}
\usepackage{bm}
\usepackage{color,soul}
\usepackage[english]{babel}
\usepackage{graphicx}
\usepackage{soul}

\begin{document}
\title{ Radiation of a Charge Moving in a Wire Structure }

\author{Sergey N. Galyamin}
\email{s.galyamin@spbu.ru}
\affiliation{Saint Petersburg State University, 7/9 Universitetskaya nab., St. Petersburg, 199034 Russia}

\author{Viktor V. Vorobev}
\affiliation{Saint Petersburg State University, 7/9 Universitetskaya nab., St. Petersburg, 199034 Russia}

\author{Andrei Benediktovitch}
\affiliation{Belarusian State University, 4 Nezavisimosti av., Minsk, 220030 Belarus}
\affiliation{CFEL, DESY, 85 Notkestrasse, Hamburg, 22607 Germany}

\date{\today}

\begin{abstract}
A theoretical approach for describing the electromagnetic radiation produced by prolonged electron bunch propagating in the lattice of metallic wires of finite length is presented.
This approach is based on vibrator antenna theory and involves approximate solving of Hallen's integral equation.
For a single wire, it is also supposed that a wire is sufficiently thin and charge motion is relativistic.
For many-wire structures, the approximation similar to kinematic approach of parametric X-ray radiation (PXR) theory is additionally applied.   
The validity of the method is verified by numerical simulations with COMSOL Multiphysics. 
Possible applications of interaction between charged particle bunches and artificial wire structures are discussed.
\end{abstract}

\maketitle

\section{Introduction\label{sec:intro}}

Artificial wire structures that have been attractive to researchers over the past several decades have not lost their importance in recent years.
One widespread motivation of mentioned researches was the ``simulation'' of plasma properties for microwave frequencies with the use of ``rodded medium'', i.e. the lattice of parallel conducting wires (rods)~%
\cite{Rot62}.
Later on, this idea was extensively utilized in the context of development of metamaterials and mainly ``left-handed'' metamaterials~%
\cite{SZKKE08}.
In the latter case, ``wire medium'' was used for providing negative effective dielectric permittivity (within certain frequency region) while negative effective magnetic permeability in this region was provided by lattice of ``split-ring resonators''~%
\cite{Pen98, Smth00, SmthKr00}.
Effective electromagnetic (EM) properties of ``wire medium'' itself was also studied in details~%
\cite{BTV02,TD11,MasSilv09}.
For example, the role of spatial dispersion and various possibilities to suppress this effect were discussed~%
\cite{DP09}.
It is worth noting that in the aforementioned investigations the wavelengths under consideration were supposed to be much larger compared to the period of wire structure, therefore allowing averaged electromagnetic description of the structure with the use of ``effective'' macroscopic parameters.  
In this ``long-wave'' framework, radiation occurring during the passage of a charged particle bunch through three-dimensional wire medium (i.e., Cherenkov radiation) or in the vicinity of planar wire structure was actively studied~%
\cite{VT12, FMS12, TV13, TV14, TVG14, TVG15, VTG17}. 
In particular, this ``long-wave'' radiation is of essential interest due to its nondivergent properties~%
\cite{VT12, FMS12}.

However, when the bunch is sufficiently short, short wavelengths (comparable with structure period) can be generated therefore altering the possibility to describe the wire structure using effective parameters.
The portion of the EM radiation related to these short enough wavelengths (``diffraction response'' or ``short-wave response'') can be described using Bragg's diffraction theory formalism, similarly to the parametric X-ray radiation (PXR) in real crystals~%
\cite{BarGurin15, BarGurn17}. 
Under the described conditions, the metallic wire assembly can be referred to as a ``wire crystal''. 
For example, if the array spacing is of order of mm, the resulting Cherenkov radiation wavelength is in THz frequency range.
Since this range is of significant interest during last decade due to its prospective applications in various areas~%
\cite{Wil06}%
, corresponding wire structures can be used for development of efficient THz radiation sources~%
\cite{AnBar15}. 
Moreover, waveguides loaded with artificial metamaterials (including specific THz wire crystals) are considered nowadays as prospective candidates for development of high-power and high-gradient accelerators~%
\cite{HAnd18,LuShapConde19}. 

In our previous paper~%
\cite{CBGVHproc18}%
, we have started developing the universal approach for investigation of EM radiation produced by a charged particle bunch moving through the wire structure composed of finite length perfectly conductive wires.
In this paper, we describe this approach in details and prove its applicability by comparison between analytical results and results of simulations with COMSOL Multiphysics.
In particular, we show that the approximation similar to the kinematic approach of PXR is applicable to the wire structure with thin enough conductors and sparse enough lattice.   

The paper is organized as follows.
After the Introduction (Sec.~%
\ref{sec:intro}%
), we formulate the problem for the EM radiation in a wire structure (Sec.~%
\ref{sec:problemstat}%
) and consider in detail the EM response of a single wire (Sec.~%
\ref{sec:1wire}%
).
Within this section, two approximations are discussed: the case where radiation influence on surface current distribution is neglected (Sec.~%
\ref{sec:local}%
) and the case with radiation corrections taken into account (Sec.~%
\ref{sec:unlocal}%
).
Section~%
\ref{sec:num}
contains numerical results, comparison between simulated and analytical results and discussion.
Section~%
\ref{sec:concl}
finishes the paper.   

\section{Problem formulation\label{sec:problemstat}}

\begin{figure}
\centering
\includegraphics[width=0.85\linewidth]{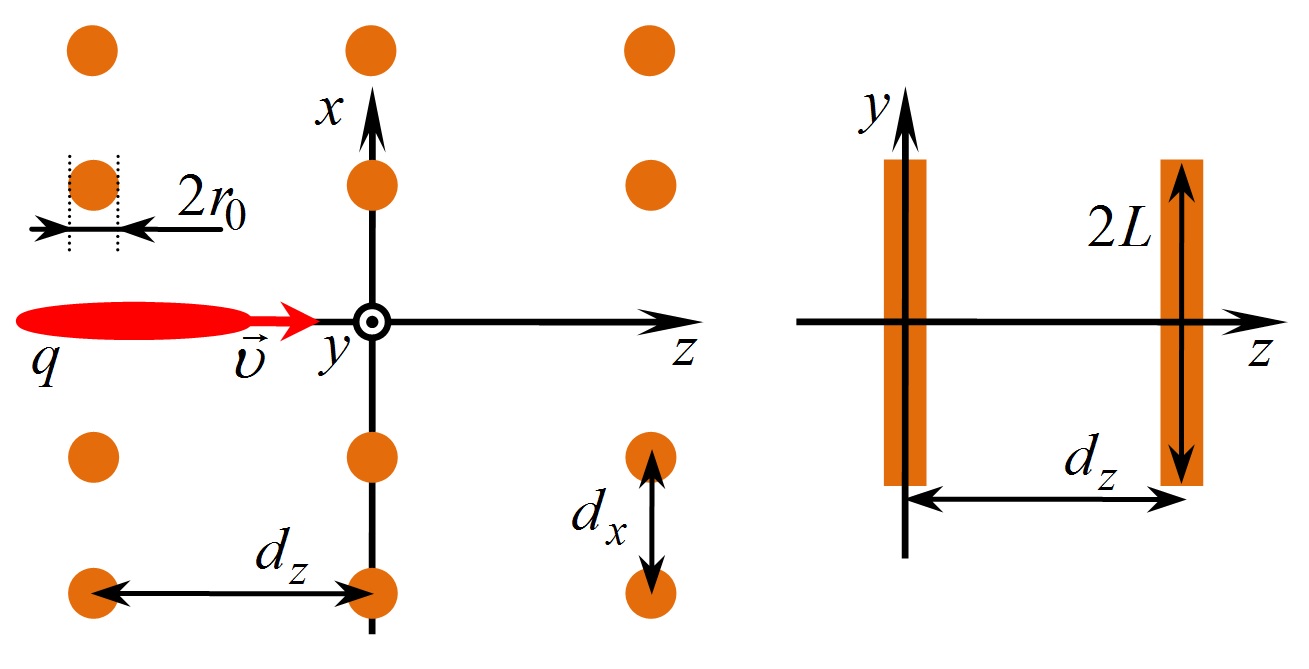}
\caption{\label{fig:geom_structure}%
Geometry of the structure: perfectly conductive wires of length
$ 2L $
and radius
$ r_0 $
form a rectangular lattice with periods
$ d_x $,
$ d_z $
which is traversed by a Gaussian bunch~
\eqref{eq:gaussian}
of charged particles.
There is no wires on bunch trajectory
$ z = 0 $.
}
\end{figure}

Figure~%
\ref{fig:geom_structure} 
shows geometry of the problem under consideration.
A thin Gaussian bunch (carrying a total charge 
$ q $ 
and having an RMS length 
$ \sigma $%
) 
with the following charge distribution
\begin{equation}
\label{eq:gaussian}
\rho ( x, y, z, t) =
\frac{ q \delta ( x ) \delta ( y ) }{ \sqrt{ 2 \pi } \sigma } 
\exp \left( \frac{ -( z - \upsilon t ) ^ 2 }{ 2 \sigma ^ 2 } \right)
\end{equation}
traverses with constant velocity
$ \upsilon =\beta c $ 
the periodic lattice of perfectly electric conductive (PEC) cylinders distributed in vacuum.
Cylinders' length is
$ 2 L $ 
and
cylinders' radius is 
$ r_0 $. 
%
%
Wires are located in nods of rectangular lattice with periods
$ d_{x} $, 
$ d_{z} $
excluding 
$ z $-%
axis along which the bunch moves.
Therefore the position of each cylinder's axis 
$ x_{ l m } $,
$ z_{ l m } $
is given by pair of integers
$(l, m)$,
so that
\begin{equation}
x_{ l m } = l d_{ x },
\quad
z_{ l m } = m d_{ z }.
\label{eq:lm}
\end{equation}
Below, we will use the following approximation: we suppose that each wire is excited by Coulomb field of the moving bunch but does not affected by the field produced by each neighboring wire.
This approximation is very close to the ``kinematic approach'' of the parametric X-ray radiation of charged particle bunches in real crystals.  
The validity of this approach will be further verified by numerical simulations in COMSOL (see Sec.~%
\ref{sec:num}%
).
The approach used below for calculation of response of each wire is related to Hallen's method widely used in antenna theory. 
Note that here we generalize this approach to the case of excitation by a charged particle bunch.
This approximation allows consideration of a single wire excitation independently. 

\section{Single wire excitation\label{sec:1wire}}

\begin{figure}
\centering
\includegraphics[width=0.85\linewidth]{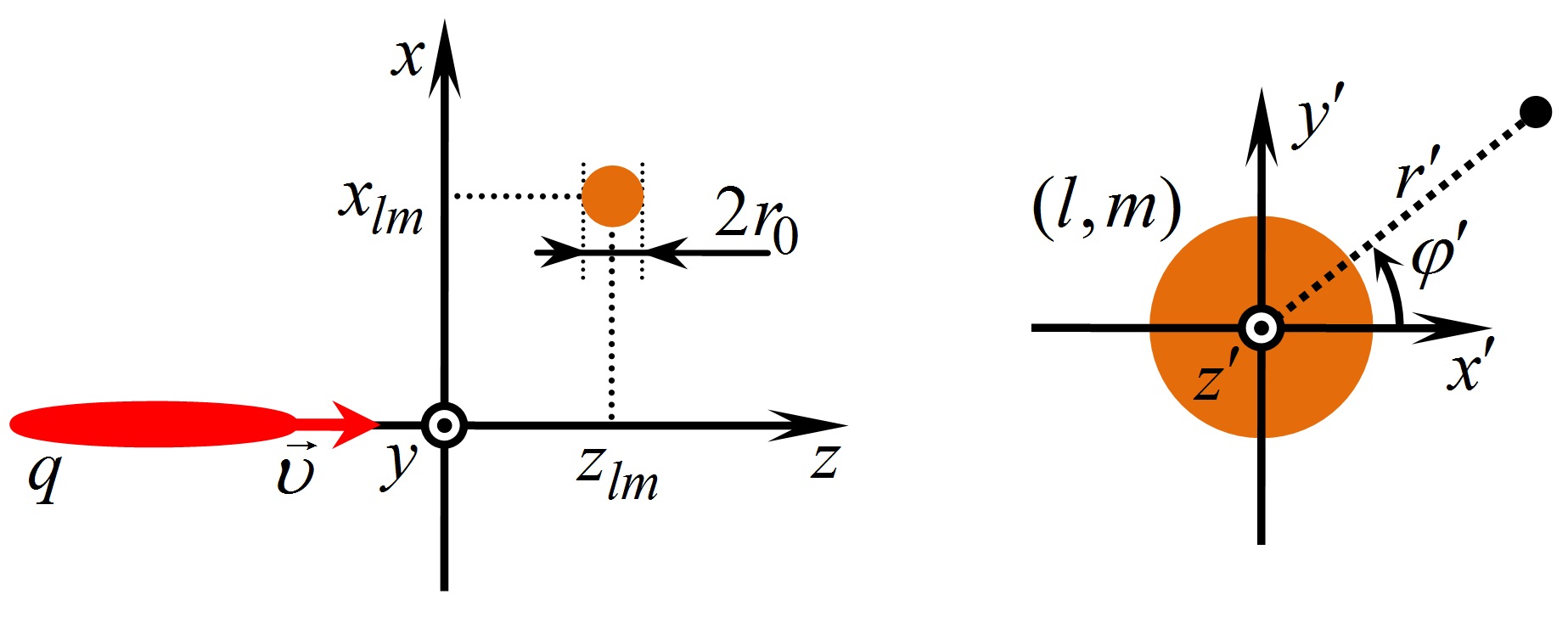}
\caption{\label{fig:geom_1wire}%
Single wire excitation and local coordinate frames (Cartesian frame
$ x^{\prime}, y^{\prime}, z^{\prime} $
and cylindrical frame 
$ r^{\prime}, \varphi^{\prime}, z^{\prime} $%
) associated with the wire.
}
\end{figure}

In this section, we will calculate the response generated by single wire with ``coordinates''
$ ( l, m ) $
aside which a Gaussian bunch moves.
The geometry of this sub-problem is shown in Fig.~%
\ref{fig:geom_1wire}.

The problem is solved in the frequency domain, therefore Fourier harmonic amplitudes are considered, i.e.
\begin{equation}
\label{eq:fourier}
\left\{
\vec{ E }_{ \omega },
\,
\vec{ H }_{ \omega }
\right\}
=
\frac{ 1 }{ 2 \pi }
\int\nolimits_{ -\infty }^{ +\infty } d t
\left\{
\vec{E},
\,
\vec{H}
\right\}
\exp{ \left( i \omega t \right) },
\end{equation}
``Incident'' field (%
$ r $%
-component in the cylindrical frame associated with the Cartesian frame 
$ x $%
,
$ y $%
,
$ z $%
) has the form: 
%
\begin{equation}
\label{eq:Erhoom}
E_{ \omega r }^{ ( i ) }
= 
\frac{ i q s_0 }{ 2 \beta c }
H_1^{ ( 1 ) } ( r s_0 )
e^{ i \frac{ \omega z }{ \upsilon } }
e^{ -\frac{ \omega^2 \sigma^2 }{ 2 \upsilon^2 } },
\end{equation}
%
%
Here 
$ r = \sqrt{ x^2 + y^2 } $, 
$ H_{ 1 }^{ ( 1 ) } $ 
is Hankel function, 
$ s_{0} = i \sigma_0 $,
$ \sigma_0 = \sqrt{ \omega^2 \upsilon^{ -2 } ( \beta^2 - 1 ) } $, 
$ \mathrm{ Re } \sqrt{ \phantom{ \sigma_0 } } > 0 $.

For 
$ \sigma = 0 $%
, expression
\eqref{eq:Erhoom}
corresponds to the field of a point charge moving in vacuum.
Below, the relativistic motion will be considered, 
$ \beta \to 1 $, 
therefore  
$ \sigma_0 \to 0 $.
Using asymptotic expressions for Hankel function~%
\cite{AS64}, 
the incident field can be simplified and 
$ E_r $
takes the form:
\begin{equation}
\label{eq:Erhoomsimp}
E_{ \omega r }^{ ( i ) }
\underset{ \beta \to 1 }
{ \mathop{ \approx } }
\frac{ q }{ \pi c } 
\frac{ 1 }{ r }
\exp \left( i \frac{ \omega z }{ c } - \frac{ \omega^2 }{ \omega_{ \sigma }^2 } \right),
\end{equation}
where
$ \omega_{\sigma } = \sqrt{2} c \beta / \sigma $. 
%
%

The main problem here is to find the surface current induced at the surface of the wire.
We will suppose that wires are thin, i.e. 
\begin{equation}
\label{eq:thinwire}
r_{0} \ll L,
\end{equation}
therefore wire flanges can be neglected.  
Moreover, we can suppose that the surface current has only 
$ y $%
-component and it does not depend on 
$ \varphi^{ \prime } $ 
(see Fig.~%
\ref{fig:geom_1wire}%
). 
Therefore, surface current
$ \vec{ j }_e^{ surf } $ 
can be presented as follows:
\begin{equation}
\label{eq:jsurf}
\vec{ j }_e^{ surf } = \vec{e}_y j_e^{ surf },
\quad
j_e^{ surf } = \frac{ I( z^{ \prime } ) }{ 2 \pi r_0 } \delta( r^{ \prime } - r_0 ),
\end{equation}
where 
$ I( z^{ \prime } )$ 
is the total current of a wire. 
The current 
$ I( z^{ \prime } )$
should satisfy the boundary conditions at the ends of a wire: 
\begin{equation}
\label{eq:Ibc}
I( L ) = I( -L ) = 0.
\end{equation}
Vector and scalar potentials satisfy the following equations:
\begin{equation}
\label{eq:Helmholtspot}
\left( 
\Delta + k_0^2 
\right)
\left\{ 
\begin{aligned}
& \vec{ A }_{ \omega } \\ 
& \Phi_{ \omega } \\ 
\end{aligned}
\right\}
=
-4 \pi 
\left\{ 
\begin{aligned}
& c^{ -1 } \vec{ j }_e^{ surf } \\ 
& \rho_{ e }^{ surf } \\ 
\end{aligned} 
\right\},
\end{equation}              
where 
$ k_0 = \omega c^{ -1 } $ 
and Lorentz gauge condition is utilized,
$ \mathrm{ div } \vec{ A }_{ \omega } - i k_0 \Phi_{ \omega } = 0$.
In accordance with~%
\eqref{eq:jsurf}%
, 
$ \vec{ A }_{ \omega } = \vec{e}_y A_{ \omega y } $, 
therefore we can conclude from Lorentz gauge condition that 
$ \Phi_{ \omega } = \frac{ 1 }{ i k_0 } \frac{ \partial A_{ \omega y } }{ \partial y } $.
For 
$ y $ 
component of electric field we obtain:
\begin{equation}
\label{eq:Ey}
E_{ \omega y }
=
-\frac{ \partial \Phi_{ \omega } }{ \partial y } + i k_0 A_{ \omega y }
=
\frac{ i }{ k_0 }
\left( 
\frac{ \partial A_{ \omega y } }{ \partial y^2 } + k_0^2 A_{ \omega y }  
\right).
\end{equation}
Solution of Eq.~%
\eqref{eq:Helmholtspot} 
for vector potential is obtained via convolution between the current and the Green function:
\begin{equation}
\label{eq:conv}
\begin{aligned}
A_{ \omega y }&( r^{ \prime }, y )
= \frac{ 1 }{ c } 
\iint\nolimits_{ S } j_e^{ surf } 
\frac{ \exp ( i k_0 R ) }{ R } d S = \\ 
&=
\frac{ 1 }{ 2 \pi c }
\int\nolimits_{ -L }^{ L } d z^{ \prime } I( z^{ \prime } ) 
\int\nolimits_{ -\pi }^{ \pi } d \varphi^{ \prime } \frac{ \exp ( i k_ 0 R ) }{ R },
\end{aligned}
\end{equation}
where 
$ S $ 
is the surface of the cylinder excluding flanges,
\begin{equation} 
\label{eq:R}
R 
= 
\sqrt{ ( r^{ \prime } )^2 + r_0^2 - 2 r^{ \prime } r_0 \cos \varphi^{ \prime } + ( y - z^{ \prime } )^2 }.
\end{equation}
Let us introduce the ``longitudinal potential'' 
$ U( y ) $, 
\begin{equation}
\label{eq:longpot}
U( y )
=
2
A_{ \omega y } ( r_0, y ),
\end{equation}
which is proportional to 
$ A_y $
calculated at the surface of the cylinder.
After a series of transformations, the connection between 
$ U( y ) $ 
and total current
$ I( y ) $
can be presented as the following integral relation:
\begin{equation}
\label{eq:longpotintegral}
U( y )
=
\frac{ 2 }{ c }
\int\nolimits_{ -L }^{ L } d z^{ \prime } I( z^{ \prime } )K( y - z^{ \prime } ),
\end{equation}
where
\begin{equation}
\label{eq:coreK}
\begin{aligned}
K( y - z^{ \prime } )
&=\frac{ 1 }{ 2 \pi }
\int\nolimits_{ -\pi }^{ \pi } d \varphi^{ \prime } \frac{ \exp ( i k_0 R_1 ) }{ R_1 }, \\
R_1
=
\left. R \right|_{ r^{ \prime } {=} r_0 }
&=
\sqrt{ 4 r_0^2 \sin^2 \left( \varphi^{ \prime } / 2 \right)
+
( y - z^{ \prime } )^2 }.
\end{aligned}
\end{equation}
Neglecting the 
$ \varphi^{ \prime } $%
-dependence in the numerator of integrand's fraction in~%
\eqref{eq:coreK}%
, we obtain
\begin{equation}
\label{eq:coreK1}
\begin{aligned}
K( y - z^{ \prime } )
&\approx
\exp \left( i k_0 \left| y - z^{ \prime } \right| \right) 
K_1( y - z^{ \prime } ), \\
K_1 ( y - z^{ \prime } )
&=
\frac{ 1 }{ 2 \pi }
\int\nolimits_{ -\pi }^{ \pi } \frac{ d \varphi^{ \prime } }{ R_1 }.
\end{aligned}
\end{equation}
where the kernel function 
$ K_1 ( y - z^{ \prime } ) $
can be expressed through the complete elliptic integral of the first kind
$ K_e ( \xi ) $~%
\cite{AS64}%
:
\begin{equation}
\label{eq:elliptic}
K_1 ( y - z^{ \prime } )
= 
\frac{ 1 }{ \pi a }
\varkappa
K_e ( \varkappa ),
\end{equation}
where
$ \varkappa = 2 r_0 / \sqrt{ ( y - z^{ \prime } )^2 + 4 r_0^2 } $. 
In particular, one can show that 
$ K_1( y - z^{ \prime } ) $ 
possesses logarithmic singularity for 
$ y \to z^{ \prime } $. 
Besides,
$ K_1( y - z^{ \prime } ) = K_1( z^{ \prime } - y ) $. 
Finally, we obtain the following integral equation connecting 
$ U $ 
and 
$ I $: 
\begin{equation}
\label{eq:longpotintegral1} 
U( y )
=
\frac{ 2 }{ c }
\int\limits_{ -L }^{ L } d z^{ \prime } I( z^{ \prime } )
\exp \left( i k_0 \left| y - z^{ \prime } \right| \right)
K_1( y - z^{ \prime } ).
\end{equation}

Now we are able to formulate the Hallen's problem for single passive wire vibrator excited by the Coulomb field of a moving charged particle bunch. 
Incident field on a wire 
$ ( l, m ) $ 
has the following form (wire thickness is neglected here):
\begin{equation}
\label{eq:Einconwire}
\begin{aligned}
\left. 
E_{ \omega y }^{ ( i ) } 
\right|_{ r^{ \prime } {=} r_0 }
&= \\
=
\left. 
E_{ \omega r }^{ ( i ) } 
\right|_{ r^{ \prime } {=} r_0 }
\frac{ z^{ \prime } }{ r_{ l m } }
&\underset{ \beta \to 1 }{ \mathop{ \approx } }
\frac{ q }{ \pi c }
\frac{ z^{ \prime } }{ r_{ l m }^{2} }
\exp \left( i \frac{ \omega z_{ l m } }{ c } - \frac{ \omega^2 }{ \omega_{ \sigma }^2 } \right),
\end{aligned}
\end{equation}
where
$ r_{ l m } = \sqrt{ x_{ l m }^2 + ( z^{ \prime } )^2 }$.
The surface current generates the following field at the surface of the wire:
\begin{equation}
\label{eq:Eyonwire}
\left. 
E_{ \omega y } 
\right|_{ r^{ \prime } = r_0 }
= 
- \frac{ 1 }{ 2 i k_0 }
\left( \frac{ d^2 U }{ d y^2 } + k_0^ 2 U \right).
\end{equation}
Standard boundary condition on the wire surface requires that total tangential electric field is zero, i.e.
\begin{equation}
\label{eq:Eybc}
\left.
\left( 
E_{ \omega y }^{ ( i ) }
{+}
E_{ \omega y }
\right)
\right|_{ r^{ \prime } = r_0 }
=
0,
\end{equation}
therefore
\begin{equation}
\label{eq:longpotODE}
\frac{ d^2 U( y ) }{ d y^2 } + k_0^2 U( y )
= 
2 i k_0 E_{ \omega y }^{ ( i ) }.
\end{equation}
Ordinary differential equation~%
\eqref{eq:longpotODE}%
, integral equation~%
\eqref{eq:longpotintegral1}
and boundary condition~%
\eqref{eq:Ibc} 
formulates the Hallen's problem for the surface current 
$ I( z^{ \prime } ) $.
 
Inhomogeneous equation~%
\eqref{eq:longpotODE} 
can be solved straightforwardly.
For simplicity, here we will consider the case of a point charge,
$ \sigma = 0 $. 
Final results can be easily transformed to the case of a Gaussian bunch.
General solution is
\begin{equation}
\label{eq:longpotgeneral}
U( y )
=
A_s \sin( k_0 y ) + 
A_c \cos( k_0 y )
+
U_{ inh },
\end{equation}
where 
$ A_s $,
$ A_c $
are constants. 
Particular solution
$ U_{ inh }( y ) $ 
can be found using method of variation of constants.
Unknown 
$ U_{ inh }( y ) $
is decomposed over known solutions [%
$\sin( k_0 y )$
and
$\cos( k_0 y )$%
] of homogeneous equation~%
\eqref{eq:longpotODE}%
: 
\begin{equation}
\label{eq:longpotvariation}
U_{ inh }( y )
=
B_s (y) \sin( k_0 y ) + B_c(y) \cos( k_0 y ),
\end{equation}                                    (3.24)
where unknown \emph{functions} 
$ B_s( y ) $ 
and
$ B_c( y ) $ 
are found from the system
\begin{equation}
\label{eq:sysvariation}
\left\{
\begin{aligned}
&\frac{ d B_s }{ d y } \sin ( k_0 y ) 
+ \frac{ d B_c }{ d y } \cos( k_0 y ) = 0, \\ 
&\frac{ d B_s }{ d y } \sin ( k_0 y) 
-\frac{ d B_c }{ d y } \cos( k_0 y ) = 2 i k_0 E_{ \omega y }^{ ( i ) }. 
\end{aligned}
\right.
\end{equation}
After a series of simple calculations we obtain: 
\begin{equation}
\label{eq:BsBc}
\begin{aligned}
B_s( y )
&=
\frac{ 2 i q \exp \left( i k_0 z_{ l m } \right) }{ \pi c }
\left[ I_c( y ) - I_c( 0 ) \right],  \\
B_c( y )
&=
\frac{ 2 q \exp \left( i k_0 z_{ l m } \right) }{ c \pi i }
\left[ I_s( y ) - I_s( 0 ) \right],
\end{aligned}
\end{equation}
where
\begin{equation}
\label{eq:IsIc}
\begin{aligned}
I_s( y )
&=
\int{ \frac{ \sin ( k_0 y^{ \prime } ) y^{ \prime } }
{ ( y^{ \prime } )^2 + x_{ l m }^2 } d y^{ \prime } }, \\
I_c( y )
&=
\int{ \frac{ \cos ( k_0 y^{ \prime } ) y^{ \prime } }
{( y^{ \prime } )^2 + x_{ l m }^2 } d y^{ \prime } }.
\end{aligned}
\end{equation}
Functions~%
\eqref{eq:IsIc}
can be expressed through elementary functions, integral sine (%
$ \mathrm{si} $%
) and integral cosine (%
$ \mathrm{Ci}$ %
) as follows~%
\cite{PBMb1}%
:
\begin{equation}
\label{eq:Ic}
\begin{aligned}
2 I_c(y)
&=
\cosh ( k_0 x_{ l m } ) \times \\
&\times 
\left[ 
\mathrm{ Ci }( k_0 y + i k_0 x_{ l m } )
+
\mathrm{ Ci }( k_0 y - i k_0 x_{ l m } ) 
\right] + \\ 
&+ i \sinh ( k_0 x_{ l m } ) \times \\
&\times 
\left[ 
\mathrm{ si } ( k_0 y + i k_0 x_{ l m } )
-
\mathrm{ si } ( k_0 y - i k_0 x_{ l m } ) 
\right], 
\end{aligned}
\end{equation}
\begin{equation}
\label{eq:Is}
\begin{aligned}
2 I_s(y)
&=\mathrm{ cosh } ( k_0 x_{ l m } ) \times \\
&\times
\left[
\mathrm{ si } ( k_0 y + i k_0 x_{ l m } ) 
+
\mathrm{ si } ( k_0 y - i k_0 x_{ l m } ) 
\right] - \\ 
&- i \mathrm{ sinh } ( k_0 x_{ l m } ) \times \\
&\times
\left[ 
\mathrm{ Ci } ( k_0 y + i k_0 x_{ l m } )
-
\mathrm{ Ci } ( k_0 y - i k_0 x_{ l m } )
\right].  
\end{aligned}
\end{equation}
where
\begin{equation}
\label{eq:siCi}
\mathrm{si}(y) = \int\limits_{ y }^{\infty} \frac{ \sin{ ( \xi ) d\xi} }{ \xi },
\quad
\mathrm{Ci}(y) = \int\limits_{ y }^{\infty} \frac{ \cos{ ( \xi ) d\xi} }{ \xi }.
\end{equation}
Constants 
$ A_s $
and
$ A_c $ 
in~%
\eqref{eq:longpotgeneral}
should be determined from the boundary condition~%
\eqref{eq:Ibc}%
.
For this, one should solve integral equation~%
\eqref{eq:longpotintegral1}.  

\subsection{\label{sec:local}Hallen's problem solution in ``quasistationary'' approximation}

In this section, we describe the simplest approximation (so called ``quasistationary'' approximation) for solution of integral equation~%
\eqref{eq:longpotintegral1}.
Let us rewrite~
\eqref{eq:longpotintegral1}
equivalently and separate out the ``local'' term:
\begin{equation}
\label{eq:locterm}
\begin{aligned}
&U ( y ) 
=
\frac{ 2 }{ c }
\int\nolimits_{ -L }^{ L } d z^{ \prime }
K_1 ( y - z^{ \prime } ) \times \\
&\times
\left[
I( y ) - 
\left( 
I( y ) - I( z^{ \prime } )
\exp \left( i k_0 \left| y - z^{\prime} \right| \right) 
\right)
\right]
= \\ 
&=
\frac{ 2 }{ c }
I( y )
\int\nolimits_{ -L }^{ L } d z^{\prime} K_1 ( y - z^{ \prime } ) - \\
&-
\frac{ 2 }{ c }
\int\nolimits_{ -L }^{ L } d z^{\prime}
\left[
I( y ) {-} I( z^{ \prime } )
\exp \left( i k_0 \left| y {-} z^{\prime} \right| \right) 
\right]
K_1 ( y {-} z^{ \prime} ). 
\end{aligned}
\end{equation}
%
%
\begin{figure}
\centering
\includegraphics[width=0.85\linewidth]{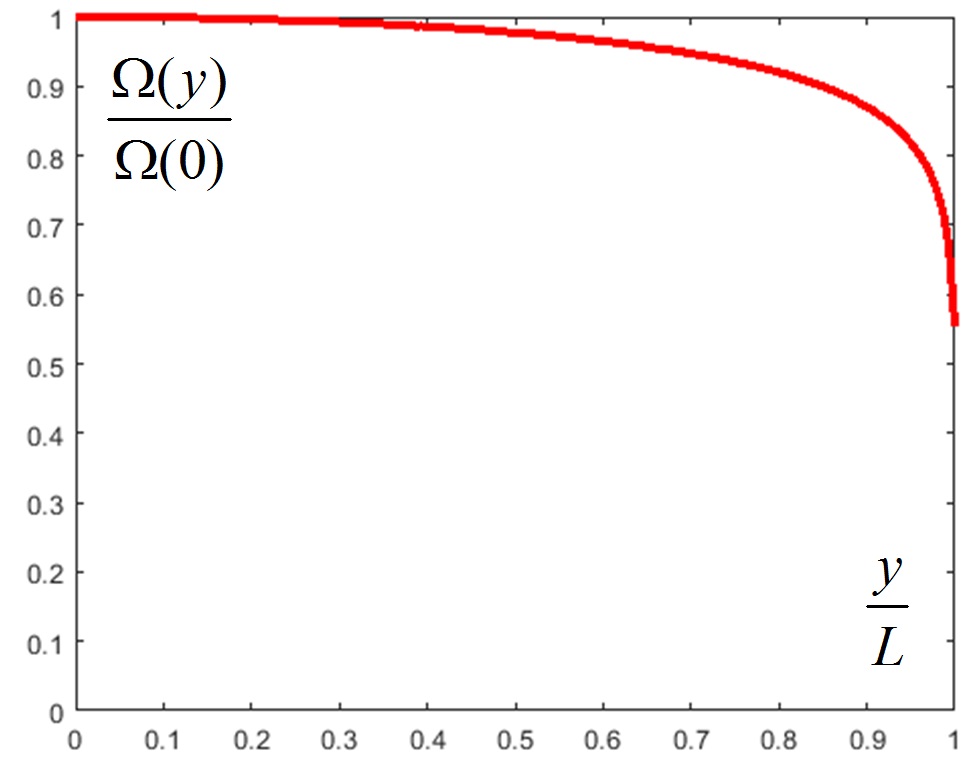}
\caption{\label{fig:Omega}%
Typical behavior of
$ \Omega( y )$ 
normalized by
$ \Omega_0 $
on the half-length of the wire for
$ L / r_0 = 200 $%
.
}
\end{figure}
%
Let us denote
\begin{equation}
\label{eq:Omega}
\Omega ( y )
=
\int\nolimits_{ -L }^{ L } d z^{ \prime } K_1 ( y - z^{ \prime } ), 
\end{equation}
therefore
$ \Omega( -y ) = \Omega( y ) $. 
Figure~%
\eqref{fig:Omega}
shows typical behavior of function 
$ \Omega( y ) $ 
for 
$ L / r_0 \gg 1 $. 
As one can see, this function changes significantly only near 
$ y = L $
therefore 
$ \Omega (y) \sim \Omega_0 $
for
$ -L < y < L $. 
For calculation of 
$ \Omega_0 $
the singular kernel 
$ K_1( y - z^{\prime} ) $ 
is frequently substituted by some regular function, for example:
\begin{equation}
\label{eq:sing2reg}
K_1 ( y - z^{\prime} ) \approx \frac{ 1 }{ \sqrt{( y - z^{ \prime } )^2 + r_0^2 } },
\end{equation}
therefore on can obtain an approximate analytical expression, 
$ \Omega_0 \approx 2 \ln ( 2 L / r_0 ) $. 
The value 
$ \Omega_0 $ 
is the large parameter of the problem, in accordance with~%
\eqref{eq:thinwire}.
However, 
$ \Omega_0 $
increases just logarithmically (i.e. relatively weakly) with an increase in ratio
$ L / r_0 $. 

Based on small parameter 
$ \alpha_0 = 1 / \Omega_0 \sim 1 / \Omega( y ) $ 
one can realize an approximate scheme for solving Eq.~%
\eqref{eq:longpotintegral1}%
.
Equivalent transformation of Eq.~%
\eqref{eq:longpotintegral1}
gives:
\begin{equation}
\begin{aligned}
U( y )
&=
\frac{ 2 }{ c } 
I( y ) \Omega( y ) 
- 
\frac{ 2 }{ c }
\int\nolimits_{ -L }^{ L } d z^{ \prime }
K_1 ( y - z^{ \prime } ) \times \\
&\times
\left[
I( y ) - I( z^{ \prime } )
\exp \left( i k_0 \left| y - z^{ \prime } \right| \right) 
\right],
\end{aligned}
\end{equation}
therefore
\begin{equation}
\label{eq:Iintegral1}
\begin{aligned}
I( y )
&=
\frac{ c }{ 2 }
\frac{ U( y ) }{ \Omega( y ) }
+
\frac{ 1 }{ \Omega( y ) }
\int\nolimits_{ -L }^{ L } d z^{ \prime }
K_1 ( y - z^{ \prime} ) \times \\ 
&\times
\left[
I( y ) - I( z^{ \prime } )
\exp \left( i k_ 0 \left| y - z^{ \prime } \right| \right) 
\right].
\end{aligned}
\end{equation}
The simplest approximation following from Eq.~%
\eqref{eq:Iintegral1}
(so called ``quasistationary approximation'') consists in neglecting the integral term. 
In such approximation we obtain the local relation between the current 
$ I $ 
and the longitudinal potential 
$ U $:
\begin{equation}
\label{eq:Ilocal}
I( y )
=
\frac{ c }{ 2 }
\frac{ U( y )}{ \Omega ( y ) }.
\end{equation}
Boundary condition
\eqref{eq:Ibc}
immediately results in 
$ U( \pm L ) = 0 $%
, therefore unknown constants in the general solution~%
\eqref{eq:longpotgeneral}
can be simply determined.
As a result, we obtain:
\begin{equation}
\label{eq:AsAc}
\begin{aligned}
A_s 
&=
\frac{ q \exp \left( i k_0 z_{ l m } \right)}
{ \pi c \sin ( k_0 L ) } \times \\
&\times
\left[
2 i I_c ( L ) \cos ( k_0 L ) - 2 I_c ( L ) \sin ( k_0 L ) 
\right], \\
A_c
&=
0.
\end{aligned}
\end{equation}
The fact that 
$ A_c = 0 $ 
is quite natural because the incident field on the wire~%
\eqref{eq:Eyonwire}
is odd with respect to
$ z^{ \prime }$%
, particular solution of the inhomogeneous equation 
$ U_{ inh } $ 
is also odd, therefore the total solution should contain odd functions only. 
Formulas~%
\eqref{eq:Ilocal},
\eqref{eq:longpotgeneral}, 
\eqref{eq:longpotvariation}, 
\eqref{eq:BsBc} 
and 
\eqref{eq:AsAc}
solve the Hallen's problem for a single wire
$ ( l, m ) $
in the approximation described above.

One important peculiarity of this approximate solution is the presence of the term
$ \sin ( k_0 L ) $
in the denominator of~%
\eqref{eq:AsAc}%
.
This term equals zero for resonant frequencies,
\begin{equation}
\label{eq:resfreq}
\omega =
\pm \omega_m = \pm 2 \pi f_m,
\,\,
f_m = \frac{ c }{ 2 L } m,
\,\,
m = 1, 2, \ldots .
\end{equation}   
For example, the first resonant wavelength 
$ \lambda_1 = c / f_1 = 2 L $%
, i.e. it equals the total length of the wire.
Therefore resonances take place when total wire length is an integer of wavelengths. 
This resonant condition will be clarified in Sec.~%
\ref{sec:unlocal}.

To conclude this subsection, we should note the following.
As it is known in antenna theory, ``quasistationary approximation'' described above does not take into account the influence of radiation on distribution of surface current along wire.
An analog of this approximation from the circuit theory is passive oscillatory circuit without active resistance (without dissipation) possessing singularity for resonant frequency.
Taking into account active resistance resolves this singularity.
For passive vibrator, taking into account the dissipation is equivalent to taking into account the radiation and it's affect on current distribution. 

\subsection{\label{sec:unlocal}Hallen's problem solution with radiation taken into account}
   
First, let us obtain solution of Eq.~%
\eqref{eq:longpotintegral1}
with higher accuracy (with respect to small parameter
$ \alpha_0 $%
) compared to the ``local'' solution~%
\eqref{eq:Ilocal}. 
To do this, we substitute into square brackets of the integrand in~%
\eqref{eq:Iintegral1}
the ``local'' solution~%
\eqref{eq:Ilocal}.
Since the integral in~%
\eqref{eq:Iintegral1}
is already multiplied by small parameter 
$ \alpha_0 $,
such a substitution is sufficient to obtain the first-order correction to the local solution~%
\eqref{eq:Ilocal}%
. 
We obtain:
\begin{equation}
\label{eq:Iunlocalsubst}
\begin{aligned}
I( y )
&=
\frac{ c }{ 2 }
\frac{ U( y ) }{ \Omega ( y ) }
+
\frac{ c }{ 2 \Omega( y ) }
\int\nolimits_{ -L }^{ L } d z^{ \prime }
K_1 ( y - z^{ \prime } )
\times \\
&\times
\left[
\frac{ U( y ) }{ \Omega( y ) }
- 
\frac{ U( z^{ \prime } ) }{ \Omega ( z^{ \prime } ) }
\exp \left( i k_0 \left| y - z^{ \prime } \right| \right) 
\right].
\end{aligned}
\end{equation}
In accordance with the form of general solution for 
$ U ( y ) $%
~%
\eqref{eq:longpotgeneral}%
, let us denote:
\begin{equation}
\label{eq:gs}
\begin{aligned}
g_s ( y )
&=
\frac{ c }{ 2 }
\frac{ \sin ( k_0 y ) } { \Omega ( y ) }
+ 
\frac{ c }{ 2 \Omega ( y ) }
\int\nolimits_{ -L }^{ L } d z^{ \prime } 
K_1 ( y - z^{ \prime } )
\times \\
&\times
\left[
\frac{ \sin( k_0 y ) }{ \Omega ( y ) }
-
\frac{ \sin( k_0 z^{ \prime } ) }{ \Omega ( z^{ \prime } ) }
\exp \left( i k_0 \left| y - z^{ \prime } \right| \right) 
\right],
\end{aligned}
\end{equation}
\begin{equation}
\label{eq:gc}
\begin{aligned}
g_c ( y )
&=
\frac{ c }{ 2 }
\frac{ \cos( k_0 y ) }{ \Omega ( y ) }
+
\frac{ c }{ 2 \Omega ( y ) }
\int\nolimits_{ -L }^{ L } d z^{ \prime } 
K_1 ( y - z^{ \prime } )
\times \\
&\times
\left[
\frac{ \cos( k_0 y ) }{ \Omega ( y ) }
-
\frac{ \cos( k_0 z^{ \prime } ) }{ \Omega ( z^{ \prime } ) }
\exp \left( i k_0 \left| y - z^{ \prime } \right| \right) 
\right],
\end{aligned}
\end{equation}
\begin{equation}
\label{eq:ginh}
\begin{aligned}
g_{ inh } ( y )
&=
\frac{ c }{ 2 }
\frac{ U_{ inh } ( y ) }{ \Omega ( y ) }
+
\frac{ c }{ 2 \Omega ( y ) }
\int\nolimits_{ -L }^{ L } d z^{ \prime }
K_1 ( y - z^{ \prime } )
\times \\
&\times
\left[ 
\frac{ U_{ inh } ( y ) }{ \Omega ( y ) }
-
\frac{ U_{ inh } ( z^{ \prime } ) }{ \Omega ( z^{ \prime } ) }
\exp \left( i k_0 \left| y - z^{ \prime } \right| \right) 
\right].
\end{aligned}
\end{equation}
Taking into account the oddness of the kernel
$ K_1 ( y - z^{ \prime } ) $
and the function
$ \Omega ( y ) $%
, one can show that:
\begin{equation}
\label{eq:gsgcginhoddness}
g_{ s, c } ( -y ) = \mp g_{ s, c } ( y ), 
\quad
g_{ inh } ( -y ) = -g_{ inh } ( y ).
\end{equation}
With these notations, total surface current is expressed as follows:
\begin{equation}
\label{eq:Iunlocalgeneral}
I ( y )
=
A_s g_s ( y )
+
A_c g_c ( y )
+
g_{ inh } ( y ),
\end{equation} 
where
$ A_s $
and
$ A_c $
are unknown constants.
Boundary conditions~%
\eqref{eq:Ibc}
lead to the following inhomogeneous linear system for 
$ A_s $
and
$ A_c $%
:
\begin{equation}
\label{eq:AsAcsystem}
\left\{
\begin{aligned}
&A_s g_s ( L ) 
+
A_c g_c ( L )
+
g_{ inh } ( L )
=0, \\ 
&A_s g_s ( -L )
+
A_c g_c ( -L )
+
g_{ inh } ( -L ) =0,
\end{aligned}
\right.
\end{equation}
with the determinant
$ D = 2 g_s ( L ) g_c ( L ) $. 
Solution of this system can be easily obtained:
\begin{equation}
\label{eq:AsAcunlocal}
A_s
=
- \frac{ g_{ inh } ( L ) }{ g_s ( L ) },
\quad
A_c = 0.
\end{equation}
Again, the fact that 
$ A_c = 0 $ 
is connected with oddness of the incident field and the function 
$ U_{ inh } $. 
Formulas~%
\eqref{eq:Iunlocalgeneral},
\eqref{eq:gs},
\eqref{eq:ginh} 
and 
\eqref{eq:AsAcunlocal} 
solve the problem.
As one can see from~%
\eqref{eq:AsAcunlocal}
and
\eqref{eq:gs}%
, singularity for resonance frequencies~
\eqref{eq:resfreq}
is absend in~%
\eqref{eq:Iunlocalgeneral}
since 
$ g_s ( L ) \ne 0 $
for
$ f = f_m $.
Therefore the obtained solution is more physical compared to the ``local'' solution obtained in Sec.~%
\ref{sec:local}%
. 

\subsection{\label{sec:scatt}Scattered electromagnetic field}

Since total surface current is found, electromagnetic field scattered by single wire can be easily calculated. 
For clarity, let us rewrite here the final expression for the current excited at the wire 
$ ( l, m ) $%
, obtained in ``quasistationary approximation'':
\begin{equation}
\label{eq:Ilocallm}
\begin{aligned}
& I^{ ( l m ) } ( z^{ \prime } )
= 
\frac{ q \exp \left( i k_0 z_{ l m } - \omega^2 / \omega_{ \sigma }^2 \right) }{ 2 \pi \Omega ( z^{ \prime } ) }
\left\{
\frac{ \sin ( k_0 z^{ \prime } ) }{ \sin ( k_0 L ) }
\right.
\times \\
&\times
\left[ 
2 i I_c ( L ) \cos ( k_0 L ) - 2 i I_s ( L ) \sin ( k_0 L ) 
\right] 
+ \\ 
&\left. 
+
2 i I_s ( z^{ \prime } )
\sin( k_0 z^{ \prime } )
-
2 i I_c ( z^{ \prime } )
\cos( k_0 z^{ \prime } ) 
\vphantom{ \frac{ q \exp \left( i k_0 z_{ l m } - \omega^2 / \omega_{ \sigma }^2 \right) }{ 2 \pi \Omega ( z^{ \prime } ) } }
\right\}.  
\end{aligned}
\end{equation} 
Here we have returned to the case of a Gaussian bunch~%
\eqref{eq:gaussian}. 
Vector potential of such a current is calculated via formulas analogous to~%
\eqref{eq:conv}:
\begin{equation}
\label{eq:Apotlm}
A_{ \omega y }^{ ( l m ) } ( x, y, z )
{ = }
\int\limits_{ -L }^{ L } d z^{ \prime }
\frac{ I^{ ( l m ) } ( z^{ \prime } ) }{ 2 \pi c } 
\int\limits_{ -\pi }^{ \pi } d \varphi^{ \prime }
\frac{ \exp \left( i k_0 R_{ 0 l m } \right) }{ R_{ l m } },
\end{equation}
where 
\begin{equation}
R_{ l m }
=
\sqrt{ ( r^{ \prime }_{ l m } )^2 + r_0^2 - 2 r_0 r^{ \prime }_{ l m } \cos \varphi^{ \prime } + ( y - z^{ \prime } )^2 },
\end{equation}
$ r^{ \prime }_{ l m } = \sqrt{ ( z - z_{ l m } )^2 + ( x - x_{ l m } )^2 } $,
and at the argument of exponent in~%
\eqref{eq:Apotlm}
we have neglected by retardation at the width of a wire, in accordance with~%
\eqref{eq:thinwire}%
, that is
\begin{equation}
\label{eq:R0lm}
\begin{aligned}
R_{ 0 l m }
&=
\left. R_{ l m } \right|_{ r_0 = 0 }
= \\
&=
\sqrt{ ( x - x_{ l m } )^2 + ( z - z_{ l m } )^2 + ( y - z^{ \prime } )^2 }.
\end{aligned}
\end{equation}
Components of electromagnetic field (which are of most interest) are calculated via standard formulas analogous to Eq.~%
\eqref{eq:Ey}%
.
In particular, we have:
\begin{equation}
\label{eq:Ex}
\begin{aligned}
E_{ \omega x }^{ ( l m ) }
&=
\frac{ i }{ k_0 }
\frac{ \partial^2 A_{ \omega y }^{ ( l m ) } }{ \partial x \partial y }
= \\
&=
\frac{ i }{ 2 \pi k_0 c }
\int\nolimits_{ -L }^{ L } d z^{ \prime }
\int\nolimits_{ -\pi }^{ \pi } d \varphi^{ \prime }
I^{ ( l m ) }( z^{ \prime } ) 
D_{ x y }^{ ( l m ) },
\end{aligned}
\end{equation}
\begin{equation}
\label{eq:Ez}
\begin{aligned}
E_{ \omega z }^{ ( l m ) }
&= \frac{ i }{ k_0 }
\frac{ \partial^2 A_{ \omega y }^{ ( l m ) } }{ \partial x \partial z }
= \\
&=
\frac{ i }{ 2 \pi k_0 c }
\int\nolimits_{ -L }^{ L } d z^{ \prime }
\int\nolimits_{ -\pi }^{ \pi } d \varphi^{ \prime }
I^{ ( l m ) } ( z^{ \prime } ) D_{ x z }^{ ( l m ) },
\end{aligned}
\end{equation}
where
\begin{widetext}
\begin{equation}
\label{eq:Dxy}
\begin{aligned}
& D_{ x y }^{ ( l m ) }
=
\frac{ \partial^2 }{ \partial x \partial y }
\frac{ \exp \left( i k_0 R_{ 0 l m } \right) }{ R_{ l m } }
= ( x- x_{ l m } ) \frac{ \exp \left( i k_0 R_{ 0 l m } \right) }{ R_{ l m } } \times  \\ 
&\times 
\left\{
( x - x_{ l m } )
\left[ 
\frac{ i k_0 }{ R_{ 0 l m } } 
- 
\frac{ r^{ \prime }_{ l m } - r_0 \cos \varphi^{ \prime } }{ r^{ \prime }_{ l m } R_{ l m }^2 } 
\right]
^2
\right.
+
\left. ( y - y^{ \prime } )
\left[ 
\frac{ - i k_0 }{ R_{ 0 l m }^3 }
+
2 \frac{ r^{ \prime }_{ l m } - r_0 \cos \varphi^{ \prime } }{ r^{ \prime }_{ l m } R_{ l m }^4 } 
\right] 
\right\}, 
\end{aligned}
\end{equation} 
\begin{equation}
\begin{aligned}
& D_{ x z }^{ ( l m ) }
=
\frac{ \partial^2 }{ \partial x \partial z}
\frac{ \exp \left( i k_0 R_{ 0 l m } \right) }{ R_{ l m } }
=
( x - x_{ l m } )( z - z_{ l m } )
\frac{ \exp \left( i k_0 R_{ 0 l m } \right) }{ R_{ l m } } 
\times  \\ 
&\times 
\left\{ 
\left[ 
\frac{ i k_0 }{ R_{ 0 l m } }
-
\frac{ r^{ \prime }_{ l m } - r_0 \cos \varphi^{ \prime } }{ r^{ \prime }_{ l m } R_{ l m }^2 } 
\right]^2 
\right.
+
\left. 
\left[ 
\frac{ -i k_0 }{ R_{ 0 l m }^3 }
-
\left( 
\frac{ 1 }{ ( r^{ \prime }_{ lm } )^2 R_{ l m }^2 }
-
\frac{ r^{ \prime }_{ l m } - r_0 \cos \varphi^{ \prime } }{ ( r^{ \prime }_{ l m } )^2 R_{ l m }^4 }
\left( 
\frac{ R_{ l m }^2 }{ r^{ \prime }_{ l m } }
+
2 r^{ \prime }_{ l m } 
\right) 
\right) 
\right] 
\right\}.  
\end{aligned}
\end{equation}
\end{widetext}
%
%
\begin{figure*}[t!]
\centering
\includegraphics[width=0.8\linewidth]{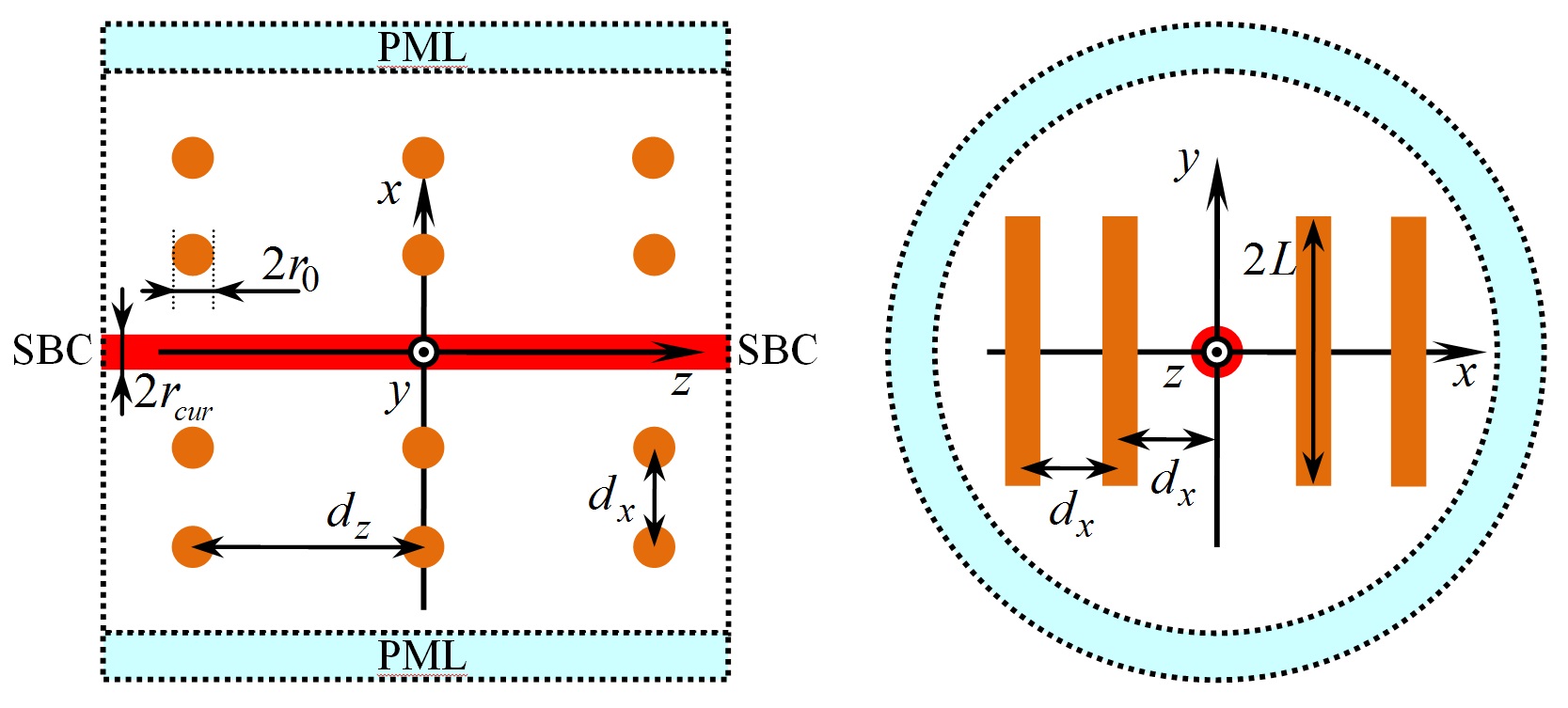}
\caption{\label{fig:comsol}%
Simulation area in COMSOL Multiphysics.
Current cylinder with radius
$r_{cur}$
(inside which the external current density
\eqref{eq:extcurdens}
is defined) is shown by red.
}
\end{figure*}
%
Let us discuss applicability of the obtained results for field calculation in wire structures containing more than 1 wire.
In our approach, it is supposed that each wire antenna is excited by the field of moving bunch only.
However, rigorously speaking, each wire in a wire structure is also excited by fields produced by all other wires. 
Therefore, the described approach can be used for many-wire structures if a response of each wire is small compared to the field of the bunch.
This should be true at least for sufficiently thin wires and sufficiently sparse lattice. 
Analogous approximation (usually refereed to as ``kinematic approach'') is widely used for consideration of parametric X-ray radiation (PXR) produced by charged particle bunches in real crystals.
The validity of the approach developed in this paper (which is also can be called ``kinematic approach'' for wire structures) will be proved and discussed in Sec.~%
\ref{sec:num}.

\section{\label{sec:num}Numerical results and discussion}

Here we present results of numerical simulations of electromagnetic processes occurring during the interaction of a charged particle bunch (or a point charge) with single wire and many-wire structures and compare this numerical results with analytical results obtained above.
For simulations, we have used frequency domain solver of COMSOL Multiphysics RF module. 
Corresponding simulated results can be directly compared with analytical ones because Fourier harmonics of the scattered field is calculated via formulas~%
\eqref{eq:Ex}%
,
\eqref{eq:Ez}
of the developed theory.
Structure of the model is shown in Fig.~%
\ref{fig:comsol}%
.
Simulation area is a cylinder with a perfectly matched layer (PML) on its back surface. 
Scattering boundary condition (SBC) was applied at the ``flanges'' of this cylinder. 
Such combination of boundary conditions allows eliminating the influence of boundaries on the results. 
In the frequency domain, a point charge moving along the axis of the structure corresponds to the threadlike current.
This current is modeled in COMSOL by traveling wave of external current density defined inside a cylinder of a small radius
$ r_{ cur } $
(``current cylinder'').
The density of this current has the form:
\begin{equation}
\label{eq:extcurdens}
\vec{ j }_{ ext }
= 
\vec{ e }_{ z }
I_{ \omega 0 }
\exp \left( i \frac{ \omega }{ c } z \right), 
\,\,
I_{ \omega 0 }
=
\frac{ q }{ 2 \pi^2 r_{ cur }^2 },
\end{equation} 
where
$ r_{ cur } \ll \lambda $ 
is radius of the current cylinder,
$ \lambda = c / f $
is wavelength under consideration.
Current~%
\eqref{eq:extcurdens}
corresponds to Fourier harmonic of the threadlike current produced by a point charge 
$ q $
moving with velocity
$ \upsilon $.
The described COMSOL model have been approved in our previous paper~%
\cite{GTV18}
Table~%
\ref{tab:comsolpars}
shows the list of parameters used for simulations. 
%
%
\begin{table}[b]
\centering
\caption{Parameters of simulations in COMSOL Multiphysics.}
\label{tab:comsolpars} 
\begin{tabular}{ c | c }
\hline
{ \bf Parameter }                    & { \bf Value }       \\
\hline
$ f $, GHz                           & $ 10 $              \\ \hline
$ \lambda $, cm                      & $ 3 $               \\ \hline
Length of the simulation area along $z$  & $ 5 c / \omega $    \\ \hline
Radius of the simulation area            & $ 4 c / \omega $    \\ \hline
$ q $, nC                            & $ 1 $               \\ \hline
$ d_x $                          & $ 0.15 c / \omega $ \\ \hline
$ d_z $                          & $ 0.15 c / \omega $ \\ \hline
$ 2 L $ (``short wire'')         & $ 0.2 c / f $       \\ \hline
$ 2 L $ (``resonant wire'')      & $ c / f $           \\ \hline
$ L / r_0 $                          & $ 200 $             \\ \hline
$ r_{ cur } $                    & $ 0.01 c / \omega $ \\ \hline
\end{tabular}
\end{table}

\begin{figure}
\centering
\includegraphics[width=0.95\linewidth]{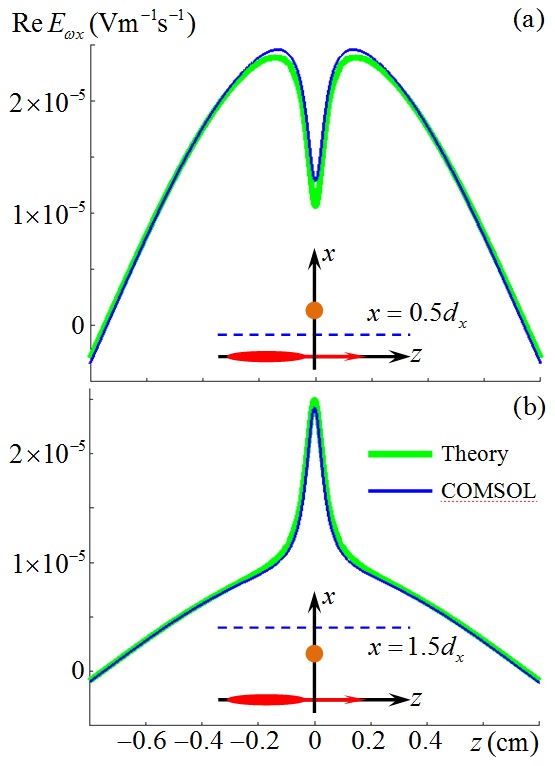}
\caption{\label{fig:1resonant}%
Real part of 
$ E_{ \omega x } $
component over
$ z $
along the line
$ x = 0.5 d_x$
(top)
and 
$ x = 1.5 d_x$
(bottom).
A spike for 
$ z = 0 $
corresponds to the response of a wire.
Wire has the ``resonant'' length, see Table~%
\ref{tab:comsolpars}%
.
}
\end{figure}

Figure~%
\ref{fig:1resonant}
shows a comparison between COMSOL and analytical results for the case of charge flight near single wire of resonant length (see Table~%
\ref{tab:comsolpars}%
) with coordinates 
$ ( 1, 0 ) $%
.

%
%

Real part of 
$ E_{ \omega x } $
component is shown on a line parallel to
$ z $%
-axis for
$ x = 0.5 d_x $%
, Fig.~%
\ref{fig:1resonant}%
(a), and for
$ x = 1.5 d_x $%
, Fig.~%
\ref{fig:1resonant}%
(b), as a function of
$ z $%
.
In accordance with
\eqref{eq:Erhoom}%
, real part of 
$ E_{ \omega x }^{ ( i ) } $ 
is proportional to
$ \cos ( \omega z / \upsilon ) $.
Background in Fig.~%
\ref{fig:1resonant}
corresponds to a half-cycle of this cosine function.
An expressed peak for
$ z = 0 $ 
corresponds to the response of a wire, as additionally illustrated by insets in Fig.~%
\ref{fig:1resonant}%
.
As one can see, the curves are in very good agreement.

%
\begin{figure}
\centering
\includegraphics[width=0.95\linewidth]{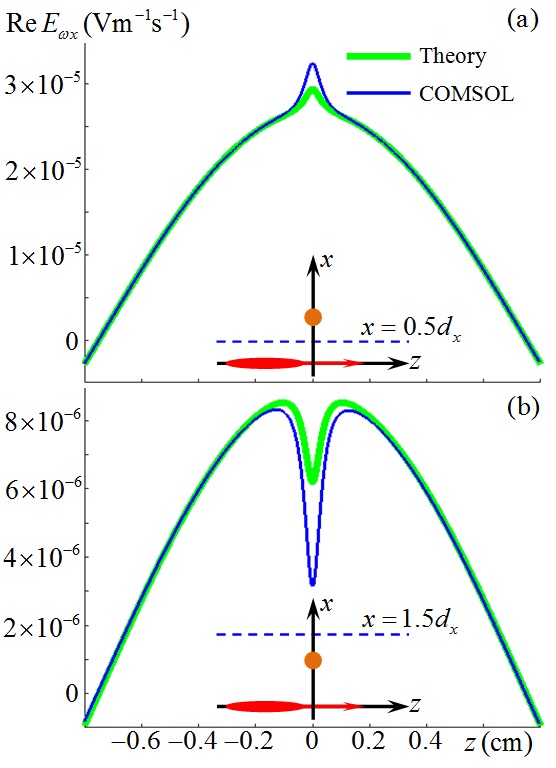}
\caption{\label{fig:1short}%
Real part of 
$ E_{ \omega x } $
component over
$ z $
along the line
$ x = 0.5 d_x$
(top)
and 
$ x = 0.5 d_x$
(bottom).
A spike for 
$ z = 0 $
corresponds to the response of a wire.
Wire is ``short'', see Table~%
\ref{tab:comsolpars}%
.
}
\end{figure}
%

%
\begin{figure}
\centering
\includegraphics[width=0.95\linewidth]{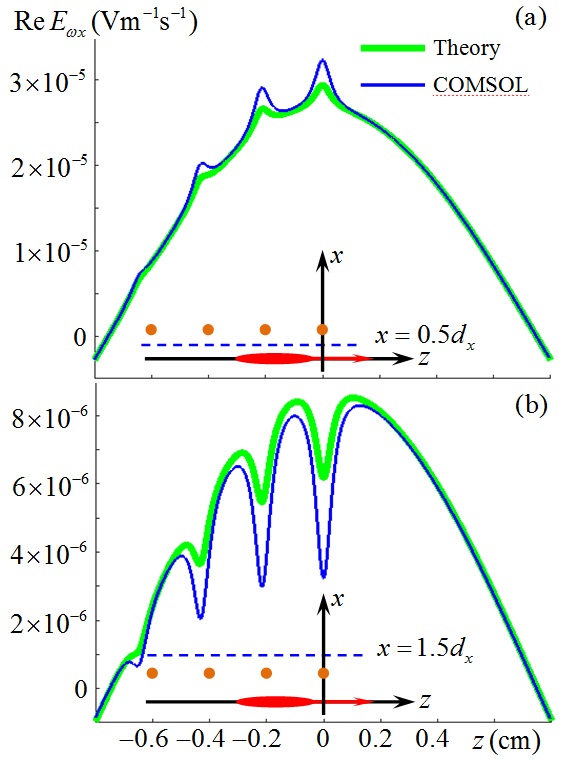}
\caption{\label{fig:4short}%
Real part of 
$ E_{ \omega x } $
component over
$ z $
along the line
$ x = 0.5 d_x$
(top)
and 
$ x = 1.5 d_x$
(bottom) for the case of 4 ``short'' wires with coordinates
$ ( 1, m ) $%
,
$ m = -3, -2, -1, 0 $%
.
}
\end{figure}
%

Figure~%
\ref{fig:1short}
shows a comparison between COMSOL and analytical results for the case of a charge flight near single ``short'' wire (see Table~%
\ref{tab:comsolpars}%
) with coordinates 
$ ( 1, 0 ) $%
.
As one can see, agreement between curves is worse, COMSOL gives stronger response compared to the theory. 
Moreover, the sign of a spike (which is responsible for wire response) has a different sign compared to the case of ``resonant wire'' (see Fig.~%
\ref{fig:1resonant}%
).
Magnitudes of spikes are also smaller compared to the ``resonant'' wire.
Figure~%
\ref{fig:4short}
shows EM field for the case of a charge flight near a string of 4 ``short'' wires with coordinates 
$ ( 1, m ) $%
, 
$ m = -3, -2, -1, 0 $%
.
As one can see, agreement between curves is similar to those in Fig.~%
\ref{fig:1short}. 
However, one can conclude from Fig.~%
\ref{fig:4short}
that each wire gives an independent response.
This fact proves the applicability of ``kinematic approach'' for wire structures used throughout the paper (each wire is excited by Coulomb field of a charge only).

One should mention that for the case of ``short'' wires we have observed just reasonable (not excellent) agreement between analytical results and COMSOL results.
Possible reason for this is that presented analytical consideration is based on approximate solution of the integral equation~%
\eqref{eq:longpotintegral1}.
In the present paper, we used this relatively simple solution to illustrate the analytical method itself.
In principle, this issue can be overcame by using numerical procedures for solving mentioned integral equation with controlled accuracy which can potentially improve the agreement between results.
However, the presented approximate solution gives surprisingly good results for a single wire of ``resonant'' length.
For many-wire structures, more sophisticated schemes taking into account the interaction between wires can be also utilized. 
In the context of using wire structures for generation of EM radiation (in particular, in THz frequency range), resonant case is of most interest since surface current magnitude (and therefore radiated EM field) is expected to be much larger in this case compared to the non-resonant wire.

\section{\label{sec:concl}Conclusion}

In the present paper, we have presented approximate analytical approach enabling calculation of EM radiation produced by a charged particle bunch during its passage through various structures composed of finite length PEC wires.
Contrary to the ``effective medium'' approach and Bragg diffraction formalism, this method is free from limitations on the relation between wavelength under consideration and structure characteristic dimensions (periods and wire length).
The method is applicable for relativistic particles and thin enough wires comprising the structure.
The limitation on wire thickness is connected with neglecting the effect of wires flanges and dependence of a surface current on azimuth angle.
Moreover, in the basis of wire thickness, it is also supposed that interaction between wires can be neglected which is closely related to the kinematic approach for PXR in real crystals. 

Presented numerical results illustrate reasonable agreement between this approach and simulations performed in RF module of COMSOL Multiphysics package.
One can conclude from this results that each wire in a many-wire structure responds independently to the Coulomb field of the bunch.
Moreover, in the resonant case (wire length equals to the wavelength) the agreement between theory and COMSOL is close to ideal (for a single wire). 

\section{Acknowledgments}

This work was supported by Russian Foundation for Basic Research (RFBR), grant No.~17-52-04107, and Belarusian Republican Foundation for Fundamental Research (BRFFR), grant No.~F17RM-026.
Numerical simulations with COMSOL Multiphysics have been performed at the Computer Center (\url{http://www.cc.spbu.ru/en}) of the Saint Petersburg State University.

%

\end{document}